\documentclass[twocolumn]{aastex62}

\definecolor{MyGreen}{rgb}{0.0,0.6,0.3}
\definecolor{MyPurple}{rgb}{0.6,0,0.3}
\usepackage{fancyref}
\usepackage{hyperref}
\hypersetup{colorlinks=true,citecolor=MyGreen,linkcolor=MyPurple,urlcolor=blue}

\begin{document}

\title[Black Hole Spins] {Most Black Holes are Born Very Slowly Rotating}

\correspondingauthor{Jim Fuller}
\email{jfuller@caltech.edu}

\author{Jim Fuller}
\affiliation{TAPIR, Mailcode 350-17, California Institute of Technology, Pasadena, CA 91125, USA}

\author{Linhao Ma}
\affiliation{TAPIR, Mailcode 350-17, California Institute of Technology, Pasadena, CA 91125, USA}
\affiliation{Department of Modern Physics, University of Science and Technology of China, Hefei, Anhui 230026, China}

\begin{abstract}

The age of gravitational wave (GW) astronomy has begun, and black hole (BH) mergers detected by LIGO are providing novel constraints on massive star evolution. A major uncertainty in stellar theory is the angular momentum (AM) transport within the star that determines its core rotation rate and the resulting BH's spin. Internal rotation rates of low-mass stars measured from asteroseismology prove that AM transport is efficient, suggesting that massive stellar cores may rotate slower than prior expectations. We investigate AM transport via the magnetic Tayler instability, which can largely explain the rotation rates of low-mass stars and white dwarfs. Implementing an updated AM transport prescription into models of high-mass stars, we compute the spins of their BH remnants. We predict that BHs born from single stars rotate very slowly, with $a \sim 10^{-2}$, regardless of initial rotation rate, possibly explaining the low $\chi_{\rm eff}$ of most BH binaries detected by LIGO thus far. A limited set of binary models suggests slow rotation for many binary scenarios as well, although homogeneous evolution and tidal spin-up of post-common envelope helium stars can create moderate or high BH spins. We make predictions for the values of $\chi_{\rm eff}$ in future LIGO events, and we discuss implications for engine-powered transients.

\end{abstract}



\section{Introduction}

Spin is one of only three fundamental properties of black holes (BHs), but there are few reliable predictions of natal black hole spins. The BH spin is determined by the angular momentum (AM) content of the core of the star that collapses into the BH. Yet our ability to predict internal stellar rotation rates and AM content has been limited by sparse observational constraints and the complex magnetohydrodynamics of differentially rotating stars. Without any AM transport within the star, nearly all compact objects would be born maximally rotating \citep{heger:00}, but efficient AM transport will couple the stellar core and envelope, slowing the spin of the core and its compact object descendant. 

Measurements of non-accreting stellar-mass BH spins are now possible for merging BHs detected by LIGO \citep{ligorun1,ligoo2:18}. Most of these BHs are consistent with very low spin \citep{roulet:18,ligoo2b:18}, though there appear to be a small fraction of moderately or rapidly rotating systems (e.g., \citealt{zackay:19}). BH spins can also be measured in X-ray binaries (XRBs), and current estimates suggest a broad range of spin-rates ($0.1 \lesssim a \lesssim 1$) \citep{miller:15}. However, XRB BH spin rates are complicated by difficult accretion disk modeling that sometimes yields conflicting results, and spins can be increased by prior/ongoing accretion \citep{fragos:15}.

Until recently, it was extremely challenging to observationally constrain AM transport within stars. Fortunately, asteroseismology has delivered decisive data \citep{beck:12,mosser:12,deheuvels:15,hermes:17,gehan:18}, unambiguously demonstrating that the internal rotation rates of low-mass stars (and their white dwarf descendants) are slower than predicted by essentially all previous models (e.g., \citealt{meynet:05,heger:05,woosley:06,cantiello:14,wheeler:15}). Most prior predictions of internal stellar rotation rates and natal NS/BH spins are therefore unreliable and could be overestimated. Models based on the Tayler-Spruit dynamo \citep{spruit:02}, such as \cite{heger:05} and \cite{qin:18}, predicted fairly slow rotation ($a \lesssim 0.1$) for BHs born from single stars, thus many BHs are likely to rotate even slower than those estimates.

In low-mass stars, AM is transported from the rapidly rotating core to the slowly rotating envelope, decreasing the spin of the stellar core and its white dwarf descendant. In a recent paper, \cite{fuller:19} demonstrated that internal rotation rates of low-mass stars can potentially be explained by magnetic torques arising from the Tayler instability (e.g., \citealt{spruit:99}), but with a different non-linear saturation mechanism than that proposed by \cite{spruit:02}, increasing AM transport and decreasing core rotation rates. Here, we extend the calculations of \cite{fuller:19} to high-mass stars to predict the AM contained in the core of the star, and hence the spin of the BH that is formed upon its collapse.

\section{Computations}

\subsection{Angular Momentum Transport}

Our stellar models include internal AM transport according to the same prescription as \cite{fuller:19} based on magnetic torques arising from the Tayler instability. These torques are larger than those predicted by the Tayerl-Spruit dynamo of \cite{spruit:02} due to a larger saturation amplitude of the Tayler instability arising from weaker non-linear damping, as elaborated in \cite{fuller:19}. In radiative zones, AM is transported by an effective viscosity
\begin{equation}
\label{AMdiff}
\nu_{\rm AM} = r^2 \Omega \bigg( \frac{\Omega}{N_{\rm eff}} \bigg)^2 \, ,
\end{equation}
where $r$ is the radial coordinate, $\Omega$ is the local angular rotation frequency, $N_{\rm eff} \approx N_\mu$ is the effective Brunt-V\"ais\"al\"a frequency, and $N_\mu$ is the compositional part of the Brunt-V\"ais\"al\"a frequency. AM is only transported via equation \ref{AMdiff} if the local shear $q = d \ln \Omega/d \ln r$ is above the critical value
\begin{equation}
\label{qmin}
q_{\rm min} = \bigg(\frac{N_{\rm eff}}{\Omega}\bigg)^{5/2} \bigg( \frac{\eta}{r^2 \Omega} \bigg)^{3/4} \, ,
\end{equation}
where $\eta$ is the magnetic diffusivity. Stellar models with this prescription provide a reasonable match with data for low-mass stars. In convective zones, AM is transported via an effective convective viscosity which enforces nearly rigid rotation.

\subsection{Stellar Models}
\label{models}

\begin{table}
\begin{tabular}{lllll}
\hline
$M_i/M_\odot$ & $Z/Z_\odot$ & $M_{\rm He}/M_\odot$ & $a_{\rm He}$ & Evolution \\
\hline
 12 & 1.2 &  3.5 & 0.006 & Single            \\
 14 & 1.2 &  4.2 & 0.007 & Single            \\
 16 & 1.2 &  4.9 & 0.007 & Single            \\
 18 & 1.2 &  5.8 & 0.008 & Single            \\
 20 & 1.2 &  6.7 & 0.009 & Single            \\
 25 & 1.2 &  9.0 & 0.009 & Single            \\
 30 & 1.2 & 10.7 & 0.010 & Single            \\
 40 & 0.5 & 16.5 & 0.003 & Single            \\
 40 & 0.1 & 19.1 & 0.014 & Single            \\
 40 & 0.01 & 21.4 & 0.010 & Single            \\
 40 & 0.5 & 13.6 & 0.050 & Case A            \\
 40 & 0.5 & 15.2 & 0.009 & Case B (stable)   \\
 40 & 0.5 & 12.3 & 0.018 & Case B (unstable) \\
 40 & 0.5 & 12.1 & 0.513 & Case B (tide)     \\
 40 & 0.012 & 31.4 & 0.549 &Homogeneous       \\
 45 & 1.2 & 17.6 & 0.010 & Single            \\
 60 & 0.5 & 26.6 & 0.006 & Single            \\
 75 & 0.5 & 35.2 & 0.035 & Single            \\
\hline
\end{tabular}
\caption{\label{spintable} Spin results for stellar models described in the text. The columns show the inital stellar mass, metallicity, final helium core mass, final helium core dimensionless spin, and the type of single/binary evolution.}
\end{table}

We construct stellar models with the MESA stellar evolution code \citep{paxton:11,paxton:13,paxton:15,paxton:18}, implementing the AM viscosity above. We study single stars with initial masses ranging from $12 \leq M_i \leq 75\,M_\odot$ and metallicities from $0.1 Z_\odot \leq Z \leq 1.2 Z_\odot$ and initial equatorial rotation speed $v_{\rm rot} = 150 \, {\rm km/s}$. All models are listed in Table \ref{spintable}. Our models include moderate convective overshoot (with exponential overshooting parameter $f=0.025$) and mass loss via the ``Dutch'' prescription (with efficiency $\eta = 0.5$). We run our models from the zero-age main sequence (ZAMS) to core carbon depletion, after which we do not expect significant changes in the helium core mass $M_{\rm He}$ or AM content $J_{\rm He}$.

\begin{figure}
\begin{center}
\includegraphics[scale=0.3]{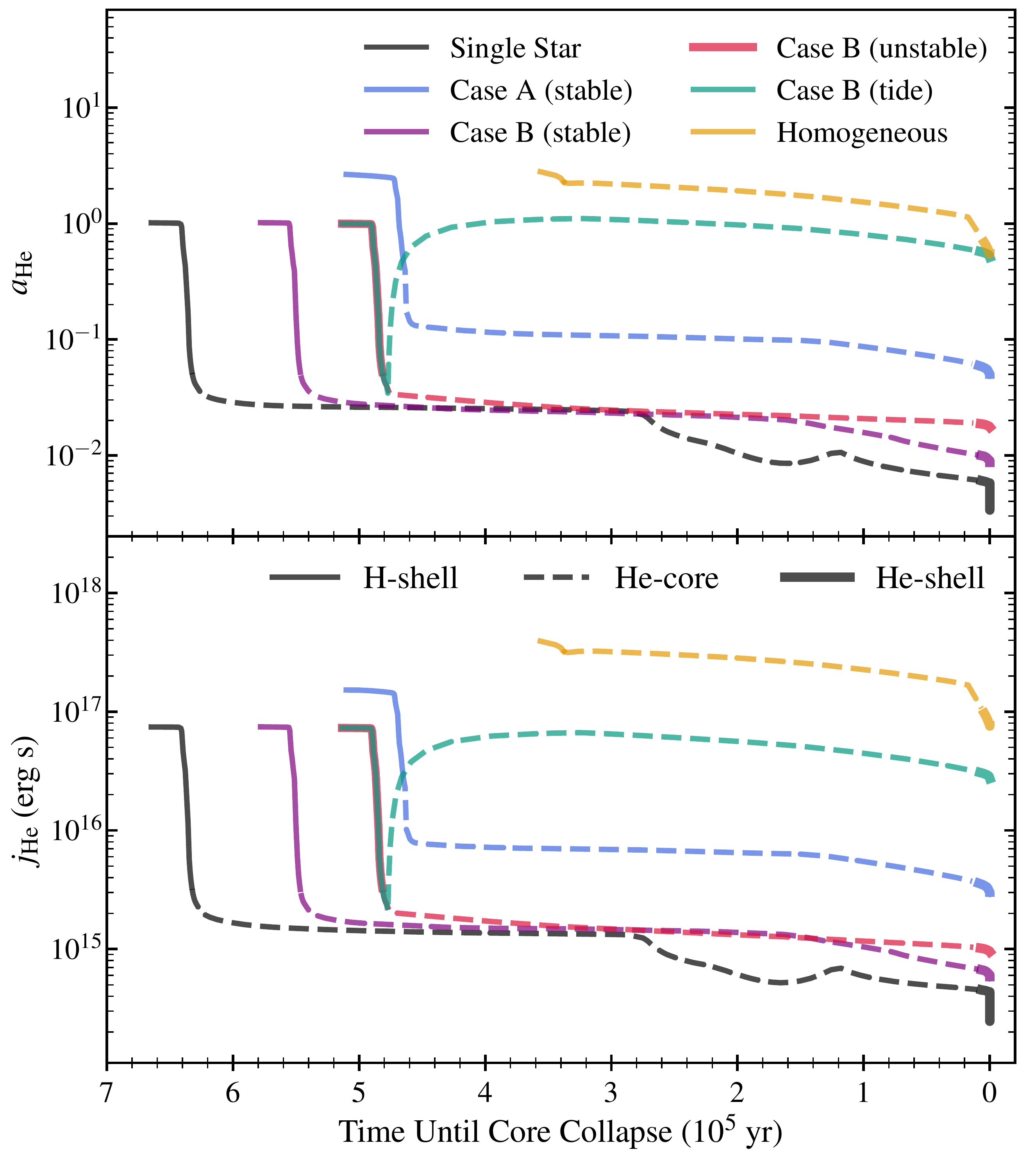}
\end{center} 
\caption{\label{DimSpinEv} {\bf Top:} Dimensionless spin $a = J_{\rm He} c/(G M_{\rm He}^2)$ of the helium core of $40 M_\odot$ progenitors as they evolve, from the end of the main sequence until carbon depletion. Each line corresponds to a single/binary scenario as discussed in the text. The line styles represent evolutionary phases corresponding to hydrogen shell-burning (solid lines), core helium-burning (dashed lines), and helium shell-burning (thick lines). If only the mass and AM of the helium core falls into the BH, the resulting spin is expected to be very small, except in a binary scenario where a helium star is tidally spun up (greed line), or a homogeneous evolutionary scenario (yellow line). {\bf Bottom:} The corresponding specific AM of the helium core, $j_{\rm He} = J_{\rm He}/M_{\rm He}$. The sudden ``cliff'' in specific AM occurs just after the main sequence, when the helium core contracts as the star crosses the Hertzprung gap.
}
\end{figure}

In addition to the single-star models listed in Table 1, we have run several binary models involving a 40 $M_\odot$ primary. In each of these models, tidal spin-up and mass transfer is included via the prescriptions of \cite{qin:18}. In the ``Case A'' scenario, the primary begins in a 3-day orbit with a companion of 20 $M_\odot$, such that mass transfer (which is assumed to be fully conservative) begins on the main sequence. In the ``Case B (stable)'' scenario, the initial orbital period is instead 50 days such that Roche lobe overflow occurs soon after the main sequence while the donor is radiative and the mass transfer is stable. In the ``Case B (unstable)'' scenario, the initial orbital period is 1000 days, Roche lobe over flow occurs when the star has expanded into a red supergiant with a convective envelope, and the mass transfer is unstable. For this model, the hydrogen envelope is removed upon Roche-lobe overflow, and the binary period is set to 3 days. The ``Case B (tide)'' scenario is the same, except the post-common envelope period is set to 0.5 days such that tides spin up the helium star. Finally, in the ``Homogeneous'' scenario, the companion mass is 40 $M_\odot$ and the initial orbital period is 1.5 days. Rotational mixing is included in this model (via MESA's default Eddington-Sweet mixing scheme) and causes the star to evolve quasi-homogeneously \citep{maeder:87,woosley:06,yoon:06,demink:09,mandel:16}.

\section{Results}
\label{sec3}

\begin{figure*}
\begin{center}
\includegraphics[scale=0.55]{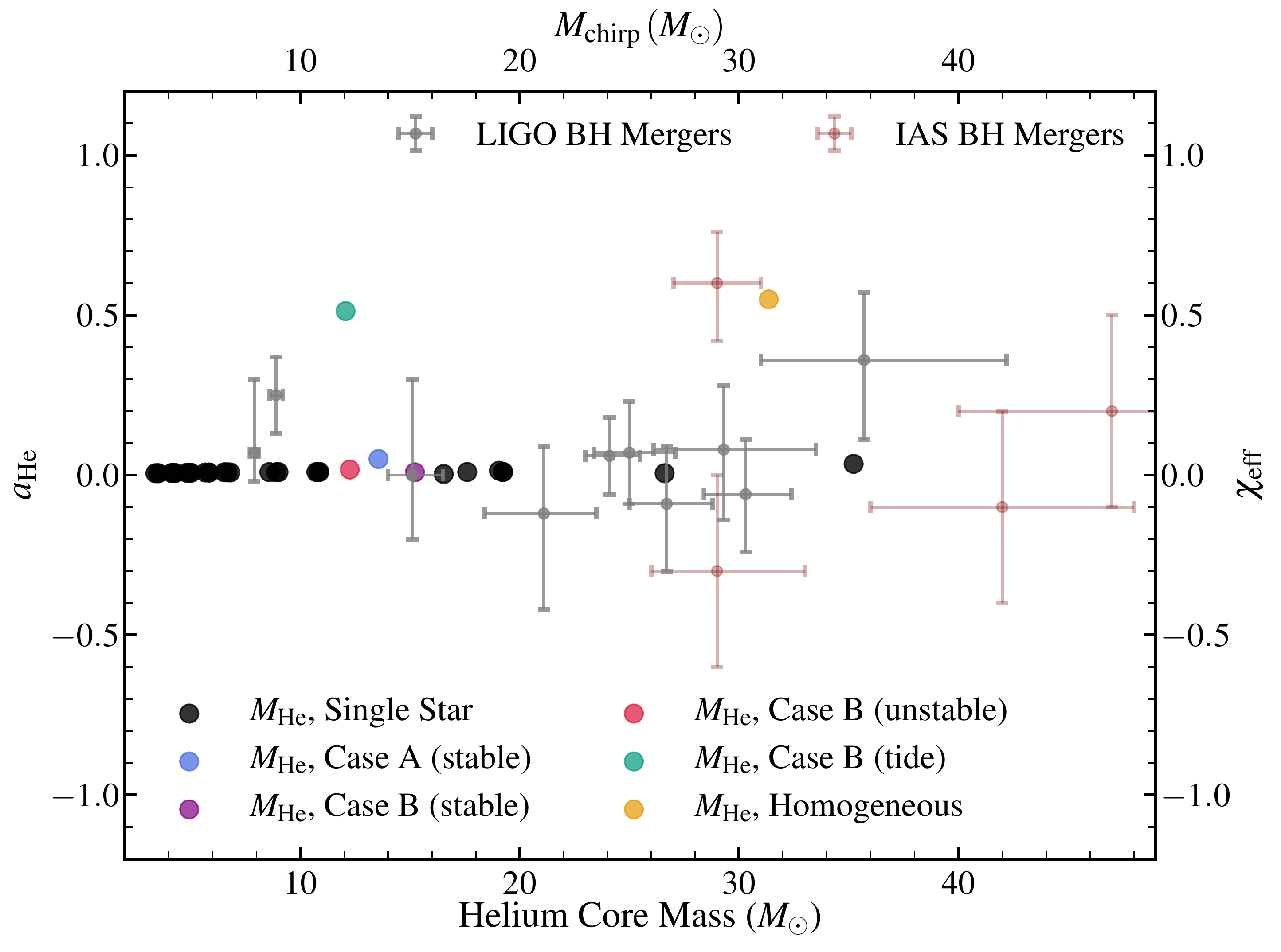}
\end{center} 
\caption{\label{DimSpin} The dimensionless spin, $a_{\rm He}$, of the helium core just before core-collapse as a function final helium core mass, with points corresponding to the models listed in Table 1. On the upper $x$-axis and right $y$-axis, we show the chirp masses and $\chi_{\rm eff}$ values for BH mergers detected by LIGO (\citealt{ligorun1,ligoo2:18}, gray crosses) and additional mergers from the IAS group (pale red crosses, \citealt{zackay:19,venumadhav:19}). For single stars (black points), if a black hole is formed upon core-collapse, we generally predict $a \sim 10^{-2}$ if only material in the helium core falls into the black hole. The colored points correspond to the same binary models shown in Figure \ref{DimSpinEv}. Only binary models with post-common envelope tidal spin-up or homogeneous evolution are capable of producing moderate or large spins.
}
\end{figure*}

Some massive stars, especially stars with initial masses $M \gtrsim 20 \, M_\odot$, produce BHs upon core-collapse. The resulting stellar remnant is a rotating Kerr BH, whose dimensionless spin $a$ is defined as 
\begin{equation}
a\equiv\frac{Jc}{GM^2} \, ,
\end{equation}
where $J$ is the AM of the BH. When core-collapse explosion fails and a BH is formed, the sudden loss of mass from radiated neutinos generates a weak shock that can still unbind the hydrodgen envelope of the star \citep{nadyozhin:80,lovegrove:13}. In red supergiants, the shock unbinds the majority of the hydrogen envelope, though blue supergiants will retain most of their hydrogen \citep{fernandez:18}. Most of our models are red supergiants or have very little remaining hydrogen at the time of collapse, so we assume that only mass within the helium core will fall into the BH. Hence, when computing BH masses and spins, we use the mass $M_{\rm He}$ and AM $J_{\rm He}$ in the helium core, which we define as the mass coordinate below which the hydrogen mass fraction falls below $10^{-2}$. 

Figure \ref{DimSpinEv} shows the dimensionless spin and specific AM of the helium core of several $40\,M_\odot$ models. When it first forms at the end the main sequence, the helium core has enough AM to produce a maximally rotating BH with $a \simeq 1$. However, similar to the results of \cite{fuller:19} for low-mass stars, the vast majority the helium core's AM is removed during hydrogen shell-burning as the helium core contracts and spins up. The internal shear activates the Tayler instability which counteracts the core spin-up, transporting AM from the helium core to the hydrogen envelope. The core's AM is further depleted by a factor of a few between helium exhaustion and core-collapse.

The final black hole spins of our single star models is typically $a \lesssim 10^{-2}$, i.e., nearly non-rotating. Figure \ref{DimSpin} shows our predictions for the dimensionless BH spin $a_{\rm He}$ for each of our models, assuming mass within the helium core collapses into a BH (though the models with $M_{\rm He} \lesssim 5 \, M_\odot$ may be more likely to form NSs). We have run models with ZAMS rotational velocities of 50, 150, and 450 km/s, but we find the initial rotation rate has almost no effect on the final value of $a_{\rm He}$, similar to low-mass stellar models. Hence, we generally predict very slow natal spins of BHs stemming from single stars near solar metallicity. A few runs at much lower metallicity also produce very slowly rotating BHs, though with slightly larger spins due to less mass loss.

Certain types of binary evolution may produce much more rapidly rotating BHs. Figure \ref{DimSpinEv} shows how the helium core AM evolves in various binary scenarios, with final spins shown as colored points in Figure \ref{DimSpin}. We predict slow BH rotation for many binaries evolving through Case A and Case B mass transfer. Even though the hydrogen envelope is eventually stripped from these models, it is still able to absorb most of the helium core's AM before it is removed, such that we still predict very slow BH rotation. There are two evolutionary scenarios that likely can result in rapid BH rotation. First, tidal spin-up of a helium star (our Case B tide model) in a short-period ($P \lesssim 1$ day) post common-envelope binary can greatly increase its AM and hence $a_{\rm He}$ (e.g., \citealt{kushnir:16}). Second, very massive, low-metallicity, and short-period binaries that evolve quasi-homogeneously never develop a core-envelope structure. The entire star is burned to helium, so the core never loses AM to an extended hydrogen envelope, allowing it to remain rapidly rotating until core-collapse to form a high-spin BH.

\section{Discussion}
\label{sec4}

The slow natal spins predicted by our models could explain the low values of the aligned spin component $\chi_{\rm eff}$ observed for most BH mergers detected by LIGO \citep{ligoo2:18}, shown in Figure \ref{DimSpin}. Indeed, several recent analyses \citep{farr:18,roulet:18,ligoo2b:18} have shown that the distribution of $\chi_{\rm eff}$ implies low spins ($a \lesssim 0.1$) if the spins of the BHs are aligned with their orbit, as expected for standard binary formation mechanisms in the field \citep{kalogera:00} unless natal BH kicks are very large. Large BH spins are disfavored even for an istropic distribution of spins as expected from BHs dynamically formed in dense stellar clusters \citep{roulet:18,ligoo2b:18}, and distributions with very low BH spins are tentatively most preferred, regardless of spin-orbit inclination \citep{ligoo2b:18}. Our results, combined with the low $\chi_{\rm eff}$ of most LIGO events, suggest that most BHs are born with low spins and that low-spin priors should be considered when analyzing LIGO data.

It may be difficult to use spin alignment to disentangle BH mergers formed via field binaries from those formed via dynamical interactions (e.g., \citealt{rodriguez:16}). If most BHs rotate very slowly, LIGO data cannot distinguish aligned and misaligned systems as expected from the field and cluster scenarios, respectively. A possible corollary of our results is that merging BHs with moderate or large $\chi_{\rm eff}$ formed from field binaries, because dynamically assembled BH binaries were not formed in tight binaries and should have very low spin. However, a caveat is the population of rapidly rotating ($a \sim 0.7$) BH primaries expected for second generation cluster mergers \citep{antonini:16,fishbach:17,gerosa:17,rodriguez:18}. Still, very high-spin mergers ($\chi_{\rm eff} \gtrsim 0.6$) are difficult to explain via second generation mergers and likely form via homogeneous evolution in which both BHs form with large spin. BH mergers with negative values of $\chi_{\rm eff}$ like GW170121 \citep{venumadhav:19} or with large misaligned spin $\chi_{\rm p}$ may form primarily via misaligned second generation cluster mergers.

The LIGO data does exhibit three events (the controversial GW151216 of \citealt{zackay:19}, the Boxing Day event GW151226, and the high-mass event GW170729) which exhibit moderate spins inconsistent with zero at 90\% confidence, though both GW151216 and GW170729 are lower significance events. Such moderate spin could be produced if one of the progenitor stars is spun up by tidal evolution and produces a rapidly rotating BH, while the other BH is slowly rotating. Indeed, our ``Case B (tide)'' point in Figure \ref{DimSpin} is similar to the measured spin of GW151226 if only the secondary is rotating such that the measured $\chi_{\rm eff}$ is reduced by a factor $M_2/(M_1+M_2)$.  A naive prediction of this tidal spin-up is that $\chi_{\rm eff}$ should exhibit a bimodal distribution with peaks at very slow spins due to binaries wide enough to avoid tidal synchronization, and moderate spins due to binaries where the second star was tidally synchronized \citep{zaldarriaga:18}. While \cite{qin:18} and \cite{bavera:19} predict a more continuous distribution, future detections will help distinguish different evolutionary pathways \citep{stevenson:17,talbot:17,farr:18,gerosa:18}. 

In these compact binaries, a weak explosion that generates a large amount of fallback material could moderately increase the BH spin above our estimates, because the fallback material is tidally torqued by the companion \citep{batta:17,schroder:18}. Alternatively, loss of mass/AM from material with enough AM to form an accretion disk around the BH could moderately decrease the BH spin \citep{batta:19}. Similar to prior works, we predict a population of moderately rotating ($\chi_{\rm eff} \sim 0.1-0.5$) BHs at a wide range of masses formed via the tidal spin-up scenario. \cite{bavera:19} predict that roughly 40\% of BH mergers detected by advanced LIGO should have $\chi_{\rm eff} > 0.1$ due to this evolutionary channel.

Forming binaries with very large spin $\chi_{\rm eff} \sim 1$ requires two aligned and rapidly rotating BHs if the mass ratio is near unity. The only model of which we are aware that can produce such events in the face of efficient AM transport is the chemically homogeneous scenario (e.g., \citealt{demink:16,mandel:16,marchant:17}). Hence, observations of $\chi_{\rm eff} \sim 1$ events may provide strong support for the homogeneous evolution scenario. The BH merger candidate GW151216 \citep{zackay:19} and GW170729 are the best candidates for homogeneous evolution thus far, and both events lie close to our ``Homogeneous'' point in Figure \ref{DimSpin}. Because the homogeneous evolution channel can only produce somewhat massive BHs, we predict an absence of highly spinning $\chi_{\rm eff} \sim 1$ and low-mass ($M_{\rm chirp} \lesssim 25 \, M_\odot$) events. Homogeneous evolution can produce either slow or moderate rotation when stellar metallicity is not small and stellar winds carry away most of the stars' AM during core helium burning. Hence at high masses ($M_{\rm chirp} \gtrsim 30 \, M_\odot$), it may be difficult to distinguish the tidal and homogeneous scenarios for moderately rotating BHs, but very large spins would be strong evidence for homogeneous evolution. While our homogeneous model resulted in a BH with $a \approx 0.5$, a model with less mass loss could yield $a>1$, and it is possible homogeneous evolution will produce a pileup of systems with $\chi_{\rm eff} \approx 1$.

Our results are in tension with the apparent high spins inferred for BHs in X-ray binaries (see \citealt{miller:15} for review). We are slightly skeptical of those model-dependent and sometimes contradictory measurements, which unfortunately cannot be calibrated against model-independent spin measurements. While the spins of BHs in low-mass X-ray binaries could be increased by accretion of AM after formation (\citealt{fragos:15}, though see also \citealt{king:99}), the spins of BHs in high-mass X-ray binaries must be natal. It is difficult to reconcile measurements of high-spin BHs in high-mass X-ray binaries with efficient AM transport \citep{qin:18b}, or with the slow spins of neutron stars \citep{miller:11}. One possibility is that a significant amount of hydrogen falls back onto BHs upon formation, increasing their spins above our estimates. However, measurements only exist for binary systems where most of the hydrogen envelope was likely stripped before core-collapse, potentially undermining the fall-back spin-up mechanism. 

Rotating blue supergiants, such as the progenitor of SN 1987A, may also give rise to rapidly rotating BHs. For these stars, neutrino-mediated mass loss will fail to unbind most of the hydrogen envelope \citep{fernandez:18}, and a rapidly rotating BH will be produced if the AM in the hydrogen envelope is accreted by the BH. However, because collapsing blue supergiants likely formed as a result of a prior binary interaction \citep{podsiadlowski:92}, few binary scenarios predict them to be the progenitors of BH mergers or X-ray binaries, though they could plausibly be progenitors of ultra-long gamma-ray bursts. However, engine-driven transients such as long gamma-ray bursts, broad-lined type Ic SNe, and superluminous type Ic SNe do not show evidence for any hydrogen in their progenitor stars. These transients are likely driven by a rapidly rotating central engine from a (mostly) carbon-oxygen progenitor star (see recent review in \citealt{fryer:19}). Our results suggest these events are unlikely to originate from single stars, except at very low metallicity ($Z \lesssim 0.004$) where homogeneous evolution can occur for single stars \citep{yoon:06}. Hence, we expect most engine-driven transients are likely produced via tidally spun-up Wolf-Rayet stars or stars evolving through homogeneous evolution.

Finally, the only competing AM transport model that may be able to explain the internal rotation rates of low-mass evolved stars is that of \cite{kissin:15}, in which stellar radiative zones rotate nearly rigidly and significant differential rotation exists in the convection zone. This model often predicts slow compact object rotation rates \citep{kissin:18}, but predicts rapid core rotation in some cases. To compare with our predictions here, future work should investigate BH rotation rates for that scenario in more detail.

\section{Conclusion}
\label{sec5}

Asteroseismic data for low-mass stars (e.g., \citealt{deheuvels:15,hermes:17,gehan:18}) has convincingly demonstrated that the cores of low-mass stars rotate at least an order of magnitude slower than predicted by most prior stellar models \citep{cantiello:14}. Previous works on massive stars (e.g., \citealt{hirschi:05,heger:05,woosley:06}) are based on physics that over-predict core rotation rates for low-mass stars, hence their predictions of compact object rotation rates are unreliable. We have re-examined BH natal spins using AM transport via magnetic torques arising from the Tayler instability \citep{spruit:99,spruit:02}, based on an updated prescription that largely matches asteroseismic data for low-mass stars and white dwarfs \citep{fuller:19}. In massive stars, we find magnetic torques extract most of the AM from the helium core just after the main sequence. 

We predict extremely slow rotation $a\sim10^{-2}$ for BHs born from single stars. We believe such AM transport is likely to be responsible for the low $\chi_{\rm eff}$ of most merging BHs detected by LIGO thus far \citep{roulet:18,ligoo2b:18}, regardless of a field binary or dynamical origin. Our preliminary investigation of BHs resulting from various binary pathways shows that very low spins are often produced in these scenarios as well. Hence, we predict that most of the LIGO BH population will be consistent with zero spin even with significantly smaller uncertainties in $\chi_{\rm eff}$. Two evolutionary scenarios leading to moderate/high BH spin are tidal torques that spin up a helium star in a short-period orbit after a common envelope event \citep{kushnir:16,qin:18}, or rapid rotation (likely enforced by tidal spin-up) and low-metallicity that allows for homogeneous evolution \citep{maeder:87,woosley:06,yoon:06}. Both scenarios can produce moderate ($a\sim 0.1-0.5$) BH spins, but only homogeneous evolution can produce very large spins with $a \sim 1$, though it should only occur for high chirp mass ($M_{\rm chirp} \gtrsim 25 \, M_\odot$) mergers.

A corollary to our results is that BH mergers with moderate or large values of $\chi_{\rm eff}$ likely originated from tidally spun-up field binaries or second generation cluster mergers. A second corollary is that gamma-ray bursts and other high-energy transients powered by rapidly rotating compact objects are likely to be formed in binaries from one of the two tidal spin-up scenarios discussed above. Future work should investigate fall-back effects, examine stars with very low metallicities, and make predictions for a general population of binaries.

\section{Acknowledgments}

We thank Will Farr, Davide Gerosa, Greg Salvesen, and Christopher Berry for enlightening conversations. This research is funded in part by an Innovator Grant from The Rose Hills Foundation and the Sloan Foundation through grant FG-2018-10515.

\bibliography{CoreRotBib}

\begin{thebibliography}{}
\expandafter\ifx\csname natexlab\endcsname\relax\def\natexlab#1{#1}\fi
\providecommand{\url}[1]{\href{#1}{#1}}

\bibitem[{{Abbott} {et~al.}(2016){Abbott}, {Abbott}, {Abbott}, {Abernathy},
  {Acernese}, {Ackley}, {Adams}, {Adams}, {Addesso}, {Adhikari}, \&
  et~al.}]{ligorun1}
{Abbott}, B.~P., {Abbott}, R., {Abbott}, T.~D., {et~al.} 2016, Physical Review
  X, 6, 041015

\bibitem[{{Antonini} \& {Rasio}(2016)}]{antonini:16}
{Antonini}, F., \& {Rasio}, F.~A. 2016, \apj, 831, 187

\bibitem[{{Batta} \& {Ramirez-Ruiz}(2019)}]{batta:19}
{Batta}, A., \& {Ramirez-Ruiz}, E. 2019, arXiv e-prints, arXiv:1904.04835

\bibitem[{{Batta} {et~al.}(2017){Batta}, {Ramirez-Ruiz}, \& {Fryer}}]{batta:17}
{Batta}, A., {Ramirez-Ruiz}, E., \& {Fryer}, C. 2017, \apjl, 846, L15

\bibitem[{{Bavera} {et~al.}(2019){Bavera}, {Fragos}, {Qin}, {Zapartas},
  {Neijssel}, {Mandel}, {Batta}, {Gaebel}, {Kimball}, \&
  {Stevenson}}]{bavera:19}
{Bavera}, S.~S., {Fragos}, T., {Qin}, Y., {et~al.} 2019, arXiv e-prints,
  arXiv:1906.12257

\bibitem[{{Beck} {et~al.}(2012){Beck}, {Montalban}, {Kallinger}, {De Ridder},
  {Aerts}, {Garc{\'{\i}}a}, {Hekker}, {Dupret}, {Mosser}, {Eggenberger},
  {Stello}, {Elsworth}, {Frandsen}, {Carrier}, {Hillen}, {Gruberbauer},
  {Christensen-Dalsgaard}, {Miglio}, {Valentini}, {Bedding}, {Kjeldsen},
  {Girouard}, {Hall}, \& {Ibrahim}}]{beck:12}
{Beck}, P.~G., {Montalban}, J., {Kallinger}, T., {et~al.} 2012, \nat, 481, 55

\bibitem[{{Cantiello} {et~al.}(2014){Cantiello}, {Mankovich}, {Bildsten},
  {Christensen-Dalsgaard}, \& {Paxton}}]{cantiello:14}
{Cantiello}, M., {Mankovich}, C., {Bildsten}, L., {Christensen-Dalsgaard}, J.,
  \& {Paxton}, B. 2014, \apj, 788, 93

\bibitem[{{de Mink} {et~al.}(2009){de Mink}, {Cantiello}, {Langer}, {Pols},
  {Brott}, \& {Yoon}}]{demink:09}
{de Mink}, S.~E., {Cantiello}, M., {Langer}, N., {et~al.} 2009, \aap, 497, 243

\bibitem[{{de Mink} \& {Mandel}(2016)}]{demink:16}
{de Mink}, S.~E., \& {Mandel}, I. 2016, \mnras, 460, 3545

\bibitem[{{Deheuvels} {et~al.}(2015){Deheuvels}, {Ballot}, {Beck}, {Mosser},
  {{\O}stensen}, {Garc{\'{\i}}a}, \& {Goupil}}]{deheuvels:15}
{Deheuvels}, S., {Ballot}, J., {Beck}, P.~G., {et~al.} 2015, \aap, 580, A96

\bibitem[{{Farr} {et~al.}(2018){Farr}, {Holz}, \& {Farr}}]{farr:18}
{Farr}, B., {Holz}, D.~E., \& {Farr}, W.~M. 2018, \apjl, 854, L9

\bibitem[{{Fern{\'a}ndez} {et~al.}(2018){Fern{\'a}ndez}, {Quataert},
  {Kashiyama}, \& {Coughlin}}]{fernandez:18}
{Fern{\'a}ndez}, R., {Quataert}, E., {Kashiyama}, K., \& {Coughlin}, E.~R.
  2018, \mnras, 476, 2366

\bibitem[{{Fishbach} {et~al.}(2017){Fishbach}, {Holz}, \& {Farr}}]{fishbach:17}
{Fishbach}, M., {Holz}, D.~E., \& {Farr}, B. 2017, \apjl, 840, L24

\bibitem[{{Fragos} \& {McClintock}(2015)}]{fragos:15}
{Fragos}, T., \& {McClintock}, J.~E. 2015, \apj, 800, 17

\bibitem[{{Fryer} {et~al.}(2019){Fryer}, {Lloyd-Ronning}, {Wollaeger},
  {Wiggins}, {Miller}, {Dolence}, {Ryan}, \& {Fields}}]{fryer:19}
{Fryer}, C.~L., {Lloyd-Ronning}, N., {Wollaeger}, R., {et~al.} 2019, arXiv
  e-prints, arXiv:1904.10008

\bibitem[{{Fuller} {et~al.}(2019){Fuller}, {Piro}, \& {Jermyn}}]{fuller:19}
{Fuller}, J., {Piro}, A.~L., \& {Jermyn}, A.~S. 2019, \mnras, 485, 3661

\bibitem[{{Gehan} {et~al.}(2018){Gehan}, {Mosser}, {Michel}, {Samadi}, \&
  {Kallinger}}]{gehan:18}
{Gehan}, C., {Mosser}, B., {Michel}, E., {Samadi}, R., \& {Kallinger}, T. 2018,
  \aap, 616, A24

\bibitem[{{Gerosa} \& {Berti}(2017)}]{gerosa:17}
{Gerosa}, D., \& {Berti}, E. 2017, \prd, 95, 124046

\bibitem[{{Gerosa} {et~al.}(2018){Gerosa}, {Berti}, {O'Shaughnessy},
  {Belczynski}, {Kesden}, {Wysocki}, \& {Gladysz}}]{gerosa:18}
{Gerosa}, D., {Berti}, E., {O'Shaughnessy}, R., {et~al.} 2018, \prd, 98, 084036

\bibitem[{{Heger} {et~al.}(2000){Heger}, {Langer}, \& {Woosley}}]{heger:00}
{Heger}, A., {Langer}, N., \& {Woosley}, S.~E. 2000, \apj, 528, 368

\bibitem[{{Heger} {et~al.}(2005){Heger}, {Woosley}, \& {Spruit}}]{heger:05}
{Heger}, A., {Woosley}, S.~E., \& {Spruit}, H.~C. 2005, \apj, 626, 350

\bibitem[{{Hermes} {et~al.}(2017){Hermes}, {G{\"a}nsicke}, {Kawaler}, {Greiss},
  {Tremblay}, {Gentile Fusillo}, {Raddi}, {Fanale}, {Bell}, {Dennihy}, {Fuchs},
  {Dunlap}, {Clemens}, {Montgomery}, {Winget}, {Chote}, {Marsh}, \&
  {Redfield}}]{hermes:17}
{Hermes}, J.~J., {G{\"a}nsicke}, B.~T., {Kawaler}, S.~D., {et~al.} 2017, \apjs,
  232, 23

\bibitem[{{Hirschi} {et~al.}(2005){Hirschi}, {Meynet}, \&
  {Maeder}}]{hirschi:05}
{Hirschi}, R., {Meynet}, G., \& {Maeder}, A. 2005, \aap, 443, 581

\bibitem[{{Kalogera}(2000)}]{kalogera:00}
{Kalogera}, V. 2000, \apj, 541, 319

\bibitem[{{King} \& {Kolb}(1999)}]{king:99}
{King}, A.~R., \& {Kolb}, U. 1999, \mnras, 305, 654

\bibitem[{{Kissin} \& {Thompson}(2015)}]{kissin:15}
{Kissin}, Y., \& {Thompson}, C. 2015, \apj, 808, 35

\bibitem[{{Kissin} \& {Thompson}(2018)}]{kissin:18}
---. 2018, \apj, 862, 111

\bibitem[{{Kushnir} {et~al.}(2016){Kushnir}, {Zaldarriaga}, {Kollmeier}, \&
  {Waldman}}]{kushnir:16}
{Kushnir}, D., {Zaldarriaga}, M., {Kollmeier}, J.~A., \& {Waldman}, R. 2016,
  \mnras, 462, 844

\bibitem[{{Lovegrove} \& {Woosley}(2013)}]{lovegrove:13}
{Lovegrove}, E., \& {Woosley}, S.~E. 2013, \apj, 769, 109

\bibitem[{{Maeder}(1987)}]{maeder:87}
{Maeder}, A. 1987, \aap, 178, 159

\bibitem[{{Mandel} \& {de Mink}(2016)}]{mandel:16}
{Mandel}, I., \& {de Mink}, S.~E. 2016, \mnras, 458, 2634

\bibitem[{{Marchant} {et~al.}(2017){Marchant}, {Langer}, {Podsiadlowski},
  {Tauris}, {de Mink}, {Mandel}, \& {Moriya}}]{marchant:17}
{Marchant}, P., {Langer}, N., {Podsiadlowski}, P., {et~al.} 2017, \aap, 604,
  A55

\bibitem[{{Meynet} \& {Maeder}(2005)}]{meynet:05}
{Meynet}, G., \& {Maeder}, A. 2005, \aap, 429, 581

\bibitem[{{Miller} {et~al.}(2011){Miller}, {Miller}, \& {Reynolds}}]{miller:11}
{Miller}, J.~M., {Miller}, M.~C., \& {Reynolds}, C.~S. 2011, \apjl, 731, L5

\bibitem[{{Miller} \& {Miller}(2015)}]{miller:15}
{Miller}, M.~C., \& {Miller}, J.~M. 2015, \physrep, 548, 1

\bibitem[{{Mosser} {et~al.}(2012){Mosser}, {Goupil}, {Belkacem}, {Marques},
  {Beck}, {Bloemen}, {De Ridder}, {Barban}, {Deheuvels}, {Elsworth}, {Hekker},
  {Kallinger}, {Ouazzani}, {Pinsonneault}, {Samadi}, {Stello}, {Garc{\'{\i}}a},
  {Klaus}, {Li}, {Mathur}, \& {Morris}}]{mosser:12}
{Mosser}, B., {Goupil}, M.~J., {Belkacem}, K., {et~al.} 2012, \aap, 548, A10

\bibitem[{{Nadezhin}(1980)}]{nadyozhin:80}
{Nadezhin}, D.~K. 1980, \apss, 69, 115

\bibitem[{{Paxton} {et~al.}(2011){Paxton}, {Bildsten}, {Dotter}, {Herwig},
  {Lesaffre}, \& {Timmes}}]{paxton:11}
{Paxton}, B., {Bildsten}, L., {Dotter}, A., {et~al.} 2011, \apjs, 192, 3

\bibitem[{{Paxton} {et~al.}(2013){Paxton}, {Cantiello}, {Arras}, {Bildsten},
  {Brown}, {Dotter}, {Mankovich}, {Montgomery}, {Stello}, {Timmes}, \&
  {Townsend}}]{paxton:13}
{Paxton}, B., {Cantiello}, M., {Arras}, P., {et~al.} 2013, \apjs, 208, 4

\bibitem[{{Paxton} {et~al.}(2015){Paxton}, {Marchant}, {Schwab}, {Bauer},
  {Bildsten}, {Cantiello}, {Dessart}, {Farmer}, {Hu}, {Langer}, {Townsend},
  {Townsley}, \& {Timmes}}]{paxton:15}
{Paxton}, B., {Marchant}, P., {Schwab}, J., {et~al.} 2015, \apjs, 220, 15

\bibitem[{{Paxton} {et~al.}(2018){Paxton}, {Schwab}, {Bauer}, {Bildsten},
  {Blinnikov}, {Duffell}, {Farmer}, {Goldberg}, {Marchant}, {Sorokina},
  {Thoul}, {Townsend}, \& {Timmes}}]{paxton:18}
{Paxton}, B., {Schwab}, J., {Bauer}, E.~B., {et~al.} 2018, \apjs, 234, 34

\bibitem[{{Podsiadlowski}(1992)}]{podsiadlowski:92}
{Podsiadlowski}, P. 1992, \pasp, 104, 717

\bibitem[{{Qin} {et~al.}(2018{\natexlab{a}}){Qin}, {Fragos}, {Meynet},
  {Andrews}, {S{\o}rensen}, \& {Song}}]{qin:18}
{Qin}, Y., {Fragos}, T., {Meynet}, G., {et~al.} 2018{\natexlab{a}}, \aap, 616,
  A28

\bibitem[{{Qin} {et~al.}(2018{\natexlab{b}}){Qin}, {Marchant}, {Fragos},
  {Meynet}, \& {Kalogera}}]{qin:18b}
{Qin}, Y., {Marchant}, P., {Fragos}, T., {Meynet}, G., \& {Kalogera}, V.
  2018{\natexlab{b}}, ArXiv e-prints, arXiv:1810.13016

\bibitem[{{Rodriguez} {et~al.}(2018){Rodriguez}, {Amaro-Seoane}, {Chatterjee},
  \& {Rasio}}]{rodriguez:18}
{Rodriguez}, C.~L., {Amaro-Seoane}, P., {Chatterjee}, S., \& {Rasio}, F.~A.
  2018, Physical Review Letters, 120, 151101

\bibitem[{{Rodriguez} {et~al.}(2016){Rodriguez}, {Zevin}, {Pankow}, {Kalogera},
  \& {Rasio}}]{rodriguez:16}
{Rodriguez}, C.~L., {Zevin}, M., {Pankow}, C., {Kalogera}, V., \& {Rasio},
  F.~A. 2016, \apjl, 832, L2

\bibitem[{{Roulet} \& {Zaldarriaga}(2018)}]{roulet:18}
{Roulet}, J., \& {Zaldarriaga}, M. 2018, ArXiv e-prints, arXiv:1806.10610

\bibitem[{{Schr{\o}der} {et~al.}(2018){Schr{\o}der}, {Batta}, \&
  {Ramirez-Ruiz}}]{schroder:18}
{Schr{\o}der}, S.~L., {Batta}, A., \& {Ramirez-Ruiz}, E. 2018, \apjl, 862, L3

\bibitem[{{Spruit}(1999)}]{spruit:99}
{Spruit}, H.~C. 1999, \aap, 349, 189

\bibitem[{{Spruit}(2002)}]{spruit:02}
---. 2002, \aap, 381, 923

\bibitem[{{Stevenson} {et~al.}(2017){Stevenson}, {Berry}, \&
  {Mandel}}]{stevenson:17}
{Stevenson}, S., {Berry}, C. P.~L., \& {Mandel}, I. 2017, \mnras, 471, 2801

\bibitem[{{Talbot} \& {Thrane}(2017)}]{talbot:17}
{Talbot}, C., \& {Thrane}, E. 2017, \prd, 96, 023012

\bibitem[{{The LIGO Scientific Collaboration} {et~al.}(2018{\natexlab{a}}){The
  LIGO Scientific Collaboration}, {the Virgo Collaboration}, {Abbott},
  {Abbott}, {Abbott}, {Abraham}, {Acernese}, {Ackley}, {Adams}, {Adhikari}, \&
  et~al.}]{ligoo2:18}
{The LIGO Scientific Collaboration}, {the Virgo Collaboration}, {Abbott},
  B.~P., {et~al.} 2018{\natexlab{a}}, arXiv e-prints, arXiv:1811.12907

\bibitem[{{The LIGO Scientific Collaboration} {et~al.}(2018{\natexlab{b}}){The
  LIGO Scientific Collaboration}, {the Virgo Collaboration}, {Abbott},
  {Abbott}, {Abbott}, {Abraham}, {Acernese}, {Ackley}, {Adams}, {Adhikari}, \&
  et~al.}]{ligoo2b:18}
---. 2018{\natexlab{b}}, arXiv e-prints, arXiv:1811.12940

\bibitem[{{Venumadhav} {et~al.}(2019){Venumadhav}, {Zackay}, {Roulet}, {Dai},
  \& {Zaldarriaga}}]{venumadhav:19}
{Venumadhav}, T., {Zackay}, B., {Roulet}, J., {Dai}, L., \& {Zaldarriaga}, M.
  2019, arXiv e-prints, arXiv:1904.07214

\bibitem[{{Wheeler} {et~al.}(2015){Wheeler}, {Kagan}, \&
  {Chatzopoulos}}]{wheeler:15}
{Wheeler}, J.~C., {Kagan}, D., \& {Chatzopoulos}, E. 2015, \apj, 799, 85

\bibitem[{Woosley \& Heger(2006)}]{woosley:06}
Woosley, S.~E., \& Heger, A. 2006, {ApJ}, 637, 914.
\newblock \url{http://dx.doi.org/10.1086/498500}

\bibitem[{Yoon {et~al.}(2006)Yoon, Langer, \& Norman}]{yoon:06}
Yoon, S.-C., Langer, N., \& Norman, C. 2006, Astronomy and Astrophysics, 460,
  199.
\newblock \url{http://dx.doi.org/10.1051/0004-6361:20065912}

\bibitem[{{Zackay} {et~al.}(2019){Zackay}, {Venumadhav}, {Dai}, {Roulet}, \&
  {Zaldarriaga}}]{zackay:19}
{Zackay}, B., {Venumadhav}, T., {Dai}, L., {Roulet}, J., \& {Zaldarriaga}, M.
  2019, arXiv e-prints, arXiv:1902.10331

\bibitem[{{Zaldarriaga} {et~al.}(2018){Zaldarriaga}, {Kushnir}, \&
  {Kollmeier}}]{zaldarriaga:18}
{Zaldarriaga}, M., {Kushnir}, D., \& {Kollmeier}, J.~A. 2018, \mnras, 473, 4174

\end{thebibliography}

\end{document}